\documentclass[12pt]{article}

\usepackage{scicite}
\usepackage{graphicx}
\usepackage{amsmath,bm}

\usepackage{times}

\newenvironment{sciabstract}{%
\begin{quote} \bf}
{\end{quote}}

\newcounter{lastnote}

\begin{document} 

\title{Mitigating Emergency Department Crowding With Stochastic Population Models}

% Place the author information here.  Please hand-code the contact
% information and notecalls; do *not* use \footnote commands.  Let the
% author contact information appear immediately below the author names
% as shown.  We would also prefer that you don't change the type-size
% settings shown here.

\author
{Gil Parnass,$^{1}$ Osnat Levtzion-Korach,$^{2}$ Renana Peres,$^{3\ast}$ Michael Assaf,$^{1\ast}$\\
\\
\normalsize{$^{1}$ Racah Institute of Physics, Hebrew University of Jerusalem, Jerusalem 91904, Israel}\\
\normalsize{$^{2}$ Shamir Medical Center, Be'er Ya'akov, Israel}\\
\normalsize{$^{3}$The Hebrew University Business school, Jerusalem, 91905, Israel}\\
\\
\small{$^\ast$To whom correspondence should be addressed:}\\
\small{E-mail: renana.peres@mail.huji.ac.il, michael.assaf@mail.huji.ac.il}
}

% Include the date command, but leave its argument blank.

\date{}
\maketitle

%%%%%%%%%%%%%%%%% END OF PREAMBLE %%%%%%%%%%%%%%%%


\baselineskip20pt



\begin{sciabstract}
Environments such as shopping malls, airports, or hospital emergency departments often experience crowding, with many people simultaneously requesting service. Crowding is highly noisy, with sudden overcrowding "spikes". Past research has either focused on average behavior or used context-specific non-generalizable models.
Here we show that a stochastic population model, previously applied to a broad range of natural phenomena, can aptly describe hospital emergency-department crowding, using data from five-year minute-by-minute emergency-department records.
The model provides reliable forecasting of the crowding distribution. Overcrowding is highly sensitive to the patient arrival-flux and length-of-stay: a 10\% increase in arrivals triples the probability of overcrowding events. Expediting patient exit-rate to shorten the typical length-of-stay by just 20 minutes (8.5\%) reduces severe overcrowding events by 50\%. Such forecasting is crucial in prevention and mitigation of breakdown events. 
Our results demonstrate that despite its high volatility, crowding follows a dynamic behavior common to many natural systems.
\end{sciabstract}

%\paragraph*{Teaser} We employ statistical physics to resolve hospital ED overcrowding which is key for healthcare management and society at large

% In setting up this template for *Science* papers, we've used both
% the \section* command and the \paragraph* command for topical
% divisions.  Which you use will of course depend on the type of paper
% you're writing.  Review Articles tend to have displayed headings, for
% which \section* is more appropriate; Research Articles, when they have
% formal topical divisions at all, tend to signal them with bold text
% that runs into the paragraph, for which \paragraph* is the right
% choice.  Either way, use the asterisk (*) modifier, as shown, to

\section*{Introduction}

We live in a crowded world. Crowded environments such as shopping malls, trains during rush hours, airports, performance venues, religious sites, and hospital-emergency-departments %(Eroglu, Machleit, and Barr 2005; Andrews, Luo, Fang and Ghose 2016;Haase, Kasper, Koch, and Müller 2019) 
\cite{eroglu2005perceived,andrews2016mobile,haase2019pilgrim}, are characterized by an influx of arriving individuals, each seeking to receive a service that is often customized to their needs, and sometimes includes clear prioritization criteria. As a result, the exit rate, and thus the number of individuals present at any given moment, greatly fluctuates across hours and days, with high volatility and sudden "spikes" of overcrowding.

Modeling crowding has been a major challenge in disciplines such as operations research, marketing, transportation research, psychology, and healthcare management \cite{ben2015association,berry2017past,cha2011association,asaro2007emergency,asaro2007impact,trotzky2021medical,kadri2014time,green2006using,wartelle2022analysis,daldoul2018stochastic,gao2014simulating,benbelkacem2019machine,chan2012optimizing,jagtenberg2015efficient}. Most studies have implemented methodologies such as queuing models \cite{green2006using,wartelle2022analysis}, econometric analysis \cite{ben2015association,berry2017past,cha2011association,asaro2007emergency,asaro2007impact,trotzky2021medical}, time series analysis \cite{kadri2014time},  dynamic choice models \cite{daldoul2018stochastic,gao2014simulating}, and performance analysis~\cite{chan2012optimizing}. 
While these methods are powerful for capturing the average behavior in the population (e.g., arrival rate, length of stay, probability of making a specific decision based on crowding etc.), they are less suitable for analyzing the fluctuating nature of crowding. Indeed, as demographic fluctuations typically scale as the square root of the population size, such models, while useful for describing large populations, become less accurate in modeling environments with smaller populations. In particular, these models do not account for outliers, bursts, and large overcrowding spikes, which characterize crowding dynamics, and at their extreme may lead to catastrophic events such as unreasonable waiting times, service breakdown, or crowd disasters.

To better capture the stochastic nature of crowding, a class of discrete event models \cite{hoot2008forecasting,cats2016dynamic,lachapelle2011mean} and machine learning algorithms \cite{benbelkacem2019machine} has been proposed. These models typically simulate a specific context (e.g., transportation choices by passengers \cite{cats2016dynamic}, hospital-emergency-department \cite{hoot2008forecasting}, and pedestrian crowds congestion \cite{lachapelle2011mean}), by decomposing it into stages, defining the inter-stage transition flow, and examining this flow subject to the parameters of the environment. Yet, lacking an analytical framework, these models provide less insight into the relative effect of each of the the model's parameters on overall crowding. Moreover, due to their "black-box" nature, these models are harder to generalize.

Here we suggest a differing approach to modeling crowded environments. We show that despite its high volatility and spiky nature, crowding can be accurately described using a simple, generalizable, analytical approach. A Langevin stochastic differential equation, describing the dynamics of the number of patients at any given moment, can capture the stochastic nature of crowded environments, including large deviations and spikes, while still enabling conducting parameter exploration and gaining key insights on the formation of crowding and the possible avenues for its mitigation.

Rooted in population dynamics, our approach connects individual-level responses with changes in population density and structure \cite{maltby2001linking}, as well as environmental variability. These models have been used to describe fluctuating population dynamics in ecology\cite{ovaskainen2010stochastic,leirs1997stochastic}, population biology \cite{mckane2004stochastic,allen2003comparison}, epidemiology \cite{mode2000stochastic,chen2005stochastic}, cell biology ~\cite{elowitz2002stochastic,kaern2005stochasticity,singh2013quantifying,
paulsson2005models,wilkinson2009stochastic,assaf2011determining}, statistical physics \cite{assaf2017wkb}, and even turbulence \cite{brown2007large,assaf2011rare,shih2016ecological}. This paper is a novel attempt to use them to address crowding effects.

We implement the model in the context of hospital emergency departments (EDs) using the complete set of records of 679,762 ED visits over five years. Mitigating ED crowding has been a top priority for health authorities and policy makers,
%(Weiss et al 2004; ACEP 2019) 
\cite{weiss2004estimating,ACEP2019Crowding} due to crowding's consequences on compromised patient care 
%(Trezciak 2003; Ben‐Yakov et al. 2015; Berry Jaeker and Tucker 2017; ACEP 2019)
\cite{trzeciak2003emergency,ben2015association,berry2017past,ACEP2019Crowding}, patient attrition %(Batt and Terwiesch 2015; ACEP 2019)
\cite{batt2015waiting,asaro2007emergency,ACEP2019Crowding}, and even higher mortality rates %(Cha et al. 2011; ACEP 2019)
\cite{cha2011association,ACEP2019Crowding}.

We find that despite its high volatility and spiky nature, ED crowding follows a dynamic behavior common to many systems in nature. The model combines analytical understanding of overcrowding mechanisms with strong forecasting capabilities and the ability to treat overcrowding spikes. It provides reliable forecasting of the average as well as the overall hourly crowding distribution. More importantly, due to its analytical nature, the model can predict how the overcrowding probabilities vary with the model parameters. This ability, absent from various discrete event and machine learning models, is an important tool in the mitigation of overcrowding and the prevention of ED breakdowns. 

\section*{A Stochastic Population Model for Crowding}
We first present the basic notations and dynamics through a mean-field deterministic model, wherein noise is ignored, and then develop the complete stochastic model. The mean-field representation is formally valid in the limit of an infinite  population.  
\subsection*{Mean-field model}

% Assuming $n_0$ initial patients, the complete time-dependent solution is:
% \begin{equation}
%     \langle n(t)\rangle = n_*+(n_0-n_*) e^{-\beta t}
% \end{equation}

Assume a service venue where individuals arrive at arrival flux denoted by $f(t)$ and exit at rate $\beta(t)$, the dynamics of the mean number of individuals $\overline{n}(t)$ in the venue reads:
\begin{equation}\label{DRE1}
\frac{d\overline{n}(t)}{dt} = f(t) - \beta(t) \,\overline{n}(t).
\end{equation}
Here, both the arrival flux and exit rate explicitly depend on time, as these constantly vary during the day. Starting with $n_0$ individuals, the solution to Eq.~(\ref{DRE1}) reads:
\begin{equation}\label{DREsol}
    \overline{n}(t) \!=\! e^{-\int_0^t \beta(s)ds}\!\int_0^t \!\!f(s) e^{\int_0^s \beta(r)dr}ds + n_0e^{-\!\int_0^t\beta(s)ds}.
\end{equation}

Solution~(\ref{DREsol}) radically simplifies by approximating $f(t)$ and $\beta(t)$ by their time-averages: $f(t)=\overline{f}$, and $\beta(t)=\overline{\beta}$. In this case, $\overline{n}(t)=n_0 e^{-\overline{\beta} t}+n_*(1-e^{-\overline{\beta}t})$, i.e., the mean number of individuals converges, after a timescale of ${\cal O}(\beta^{-1})$, to the stable fixed point at  $n_* = \overline{f}/\overline{\beta}$.

\subsection*{Stochastic model} 
In actual crowded environments, besides their deterministic variations, the arrival flux and exit rates contain a stochastic component. We therefore incorporate two types of noise into the mean-field dynamics: inter-individual, and systematic. The inter-individual noise emanates from heterogeneity in the arrival flux, discreteness of individuals, or the type of service sought by each individual, and is sometimes termed
”demographic”, or ”internal”. The systematic noise is caused by changes in the facility infrastructures, variations in the number and quality of staff, and inconsistencies in organizational procedures. This noise is sometimes termed ”external”.
Consequently, the deterministic rate equation~(\ref{DRE1}) gives way to a stochastic differential equation for the momentary number of individual present in the venue. Using the Langevin notation~\cite{gardiner1985handbook}, we have: \begin{eqnarray}\label{Lang}
\frac{dn(t)}{dt} =   f(t)\!+\!\sqrt{f(t)\!+\!\beta(t)n(t)}\xi_1(t)-\beta(t)\left[1+\xi_2(t)\right]n(t).
\end{eqnarray}
This equation includes a deterministic term $f(t)-\beta(t)n(t)$, identical to Eq.~(\ref{DRE1}). In addition there are two noise terms. The first, $\sqrt{f(t)\!+\!\beta(t)n(t)}\xi_1(t)$, represents the inter-individual noise \cite{assaf2010extinction,assaf2017wkb}, and emanates from the underlying continuous-time master equation, [see SI, Appendix A]. The second term, $\beta(t)\xi_2(t)n(t)$, corresponds to  the systematic noise; it is multiplicative and scales with the population size~\cite{lande2003stochastic}, as it equally influences all individuals. We assume that $\xi_1(t)$ and $\xi_2(t)$ are mutually independent, zero-mean, delta-correlated (in time) noise terms with  magnitudes $\sigma_1$ and $\sigma_2$, respectively, such that $\langle\xi_i(t)\rangle = 0$, and $\langle\xi_i(t)\xi_i(t+\tau)\rangle = \sigma^2_i\delta(\tau)$, where $\delta(\tau)$ is the Dirac delta function. Notably, we have confirmed that taking $\xi_i$ with finite correlation time (i.e., colored noise, when $\xi_i$ satisfies, e.g., an Ornstein-Uhlenbeck equation~\cite{gardiner1985handbook}), does not qualitatively change the model's results [see SI, Appendix B]. Here, it is important to note that,  a qualitatively similar equation to Eq.~(\ref{Lang}) can in principal be obtained using methods from queuing theory, by taking a processor-sharing server with capacity scaling linearly with the population size $n$, and adding external noise to the capacity~\cite{kleinrock1967time}. 

A histogram over the different realizations obtained by simulating  Langevin equation~(\ref{Lang})  provides the complete statistics of events  including extreme overcrowding. This histogram can be found by transforming Eq.~(\ref{Lang})  into a Fokker-Planck equation~\cite{gardiner1985handbook} (see SI, Appendix A). The latter describes the dynamics of $P(n,t)$ -- the probability distribution of observing $n$ individuals at time $t$ in the venue -- and provides important insights into the characteristics of $P(n,t)$ and its moments.

\section*{ED Data and Crowding Metrics}
\begin{figure}[htb!]
\centering
\includegraphics[width=0.90\linewidth]{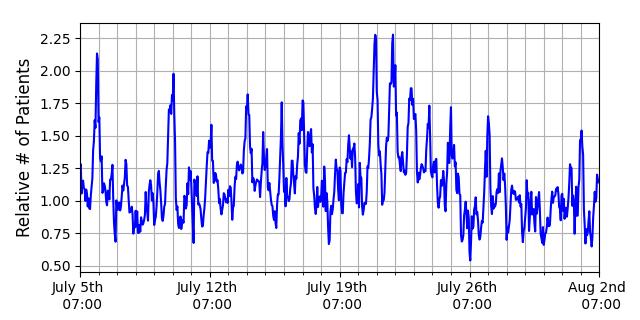}
\caption{
Relative number of patients (number of patients divided by their hourly mean) as function of time for a typical month of the data (July 5th until August 2nd, 2015).
}
\label{fig:fig1}
\end{figure}

We assembled a unique dataset containing the complete records of 679,762 visits by 333,471 unique patients, which are all the ED visits between January 1, 2013 and June 30, 2018, at a large state-run Israeli hospital. The data include the time of arrival to the ED and the time of departure (discharge or hospitalization), as well as time stamps for each recorded procedure, and many other additional variables [see Methods]. To align with the ED shift structure, we grouped the hours, when needed, into morning  (7:00-15:00), afternoon  (15:00-23:00), and night  (23:00-7:00) shifts. To demonstrate the spiky nature of the data, we show in Fig.~\ref{fig:fig1} the momentary number of patients normalized by their hourly mean, for a typical period of 4-weeks. During this period one has $\sim 15-16$ spikes exceeding the mean by $50\%$, and $\sim 4$ spikes exceeding the mean by $100\%$, indicating a quite significant noise level.

\begin{figure}[htb!]
\centering
\includegraphics[width=0.90\linewidth]{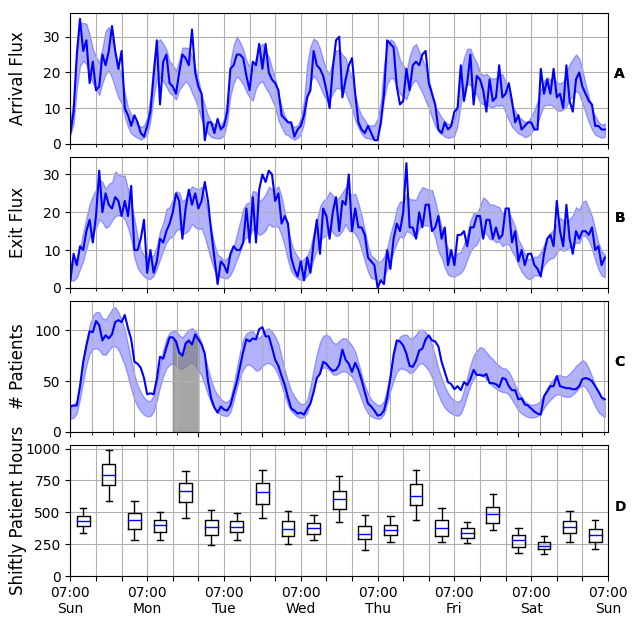}
\caption{Timeline of the arrival flux (A), exit flux (B), and momentary number of patients in the ED (C). The solid line is the hourly value for a typical week (July 5-12, 2015); the shaded blue area denotes one standard deviation from the average over all weeks. The grey area in (C) denotes the value of the patient hours for the Monday afternoon shift.
(D) Distribution of patient hours per shift. The box extends from the 25th to the 75th percentile of the shift data (whiskers mark the 5th and 95th percentile), with a line at the median. 
A week begins at 7:00 on Sunday, while the vertical grid-lines represent the shifts (07:00, 15:00 and 23:00).}
\label{fig:fig2}
\end{figure}

We measure four metrics: arrival flux, exit flux, momentary number of patients, and ``patient hours''.
The arrival flux $f(t)$ is the hourly number of incoming patients (Fig.~\ref{fig:fig2}A). The exit flux is the number of patients who left the ED per hour (for both discharge and hospitalization, see Fig.~\ref{fig:fig2}B). The momentary number of patients is the number of patients who are currently in the ED, denoted by $n(t)$ (Fig.~\ref{fig:fig2}C). Finally, the metric of patient hours is defined by $C =\int_{T}^{T+\Delta T}n(t)dt$. This is the accumulated number of patients in the ED during the time interval $[T,T+\Delta T]$, where T is measured in hours. Figure~\ref{fig:fig2}D shows the distribution of patient hours over weekly shifts. The data show daily and weekly cycles, with considerable fluctuations. 

\section*{Results - Estimate Crowding}
\begin{figure}[htb!]
\centering
\includegraphics[width=0.90\linewidth]{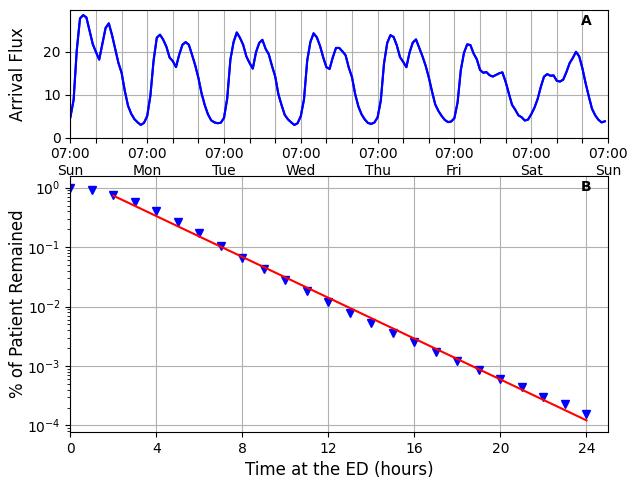}
\caption{
The functional shape of $f(t)$ and $\beta(t)$. (A) The average hourly arrival flux over a week. (B) The probability of a patient to remain in the ED since arrival versus time  (a semi-logarithmic plot). The red line is an exponential approximation, indicating that the patients' exit process is a Poisson process, with an expected value that equals the patients' average length of stay.
}
\label{fig:fig3}
\end{figure}

The model in Eq.~(\ref{Lang}) requires the estimation of the arrival flux $f(t)$, the exit rate $\beta(t)$, and the noise magnitude parameters $\sigma_1$ and $\sigma_2$.  We take $f(t)$ as the empirical arrival flux, averaged over the various weeks (Fig.~\ref{fig:fig3}A). Notably, other theoretical choices of the arrival flux are possible, for example, by fitting the data to a trapezoid function of time for each day. We confirmed (see SI, Appendix C) that this choice provides results with comparable accuracy to those shown below. For the exit rate $\beta(t)$, the data indicate that individuals exit the ED in a Poisson manner, i.e., the probability of not exiting until time $t$ is given by $p(t)=e^{-\beta t}$ (Fig.~\ref{fig:fig3}B). That is, the individual exit rate can be regarded as a constant number $\beta$ that depends on the specific day of the week (altogether four $\beta$ parameters: Sunday, midweek (Monday-Thursday), Friday, Saturday). 

We estimate the model parameters in two steps: First, we estimate the $\beta$ parameters using solution~(\ref{DREsol}) to the mean field equation.~(\ref{DRE1}) for the constant-$\beta$ case. 
Figure~\ref{fig:fig4} shows the model's results for the mean exit rate (Fig.~\ref{fig:fig4}A), mean number of patients (Fig.~\ref{fig:fig4}B) and mean patient hours (Fig.~\ref{fig:fig4}C) compared with empirical data. A high level of fit in all metrics (see Methods for goodness-of-fit statistics) can be seen. In the second step, we used the values of $\beta$, and applied maximum likelihood estimation to fit the noise parameters, $\sigma_1$ and $\sigma_2$ [see Methods].
%how taking the empirical flux from data and a constant exit rate (depending on the weekday) gives way to an accurate estimation of the mean number of patients. 
%We first estimate the average number of patients as a function of time. Figure \ref{fig:fig4} illustrates the results of the average arrival flux (\ref{fig:fig4}.A), average hourly exit flux (\ref{fig:fig4}.B), and the average momentary number of patients in the ED (\ref{fig:fig4}.C) for the model (see also Fig. S1, which incorporates additional modeling choices). Fit levels are high [see Methods section]
%Interestingly, most of the difference stems from the off-peak hours (nights and early mornings). 
%%

The estimated model enables prediction of the statistics of crowding events in the ED. 
Figure~\ref{fig:fig5}A shows excellent agreement between the model and the data with respect to the standard deviation of the number of patients over the week (see Methods for goodness-of-fit statistics). Furthermore, Fig.~\ref{fig:fig5}B demonstrates that the prediction of the model for the entire patient-number distribution  over a given ED shift agrees well with the distribution in the data (the inset shows the cumulative distribution). 
% For example, the model predicts that the probability to 55 or more patients in the ED in any shift (55 is the hourly average number of patients in the ED) is, on average $0.515\pm 0.001$, compared with $0.491$ in the data. 
% {\color{red} The probability to observe 120 patients or more in the ED in any shift (120 is defined in by the ED management as severe overcrowding), is, on average $0.79\%\pm0.04\%$, compared with $1\%$ in the data. This probability is equivalent to 5.7 hours per month.  Maybe change for 140?}
% {color{red}High agreement with the data is also obtained when computing the cumulative distribution function for \emph{relative} crowding, $x = C_{shift}/\langle C\rangle$, measured in the patient hours during a shift relative to the shift average (Fig.~\ref{fig:fig5}C). For example, roughly half of the shifts are above the average measured crowding (53\% of the data and 45\% of the model). Detailed goodness of fit measures are provided in the Methods section. Change for better data point, 20\% for example}
For example, the model predicts that the probability of observing between 80 and 88  patients in the ED in the 10 most crowded shifts is $0.119\pm 0.007$, compared to $0.120$ in the data. 
 % in the shift = $0.152\%\pm0.003\%$, compared with $0.162\%$ in the data. This probability is equivalent to 6.7 hours per year.
 Very good agreement with the data is also obtained when computing the distribution function for \emph{relative} crowding, $x = C_{shift}/\langle C\rangle$, measured in the patient hours during a shift relative to the shift average (Fig.~\ref{fig:fig5}C). For example, the probability to observe relative crowding between 1 and 1.1 (between average and 10\% over average) is $24.4\%\pm0.1\%$ in the model compared to $24.0\%$ of the data.
% Note, that in Fig.~\ref{fig:fig5} (see also Fig.~\ref{fig:Pn_vs_n_AB}) the PND on the left is unimodal whereas the PND on the right is bimodal. The reason for that is that the 5 busiest shifts are highly crowded during the entire shift, while the 5 next busiest shifts include a steady rise of the average number of patients, which results in a bimodal distribution.

\subsection*{Estimating overcrowding events}
The ability to accurately predict the probability of overcrowding is an important tool for anticipating and mitigating ED breakdowns. Overcrowding can occur when (i) the number of patients in the ED exceeds a certain absolute threshold, or (ii) when a large relative deviation above the average crowding  occurs at a given time. While the absolute numbers indicate the load relative to the existing infrastructures (e.g., beds, physical capacity, staff availability), the relative definition is indicative of the subjective perception of crowding. 20 patients arriving unexpectedly to the ED can cause overcrowding if arriving over the weekend, but can be easily processed during weekdays, when the ED is usually prepared for many arrivals. 

We follow the classification of NEDOCS \cite{weiss2004estimating}, which ranks crowding levels on a scale from 1 to 6, where the top two levels are severe  and dangerous overcrowding, which are respectively equivalent in our data to $\sim\!120$ and $\sim\!140$ patients. 
%In the data, the probability for severe and dangerous overcrowding (i.e., the probability to exceed 120 patients) is given by $1.0\%$, namely $\sim\!7.2$ hours a month, whereas the model predicts a probability of $0.79\%$, equivalent to $\sim\!5.7$ monthly hours, see Fig.~\ref{fig:fig5}B. 
For example, the model's prediction of the probability of observing  dangerous overcrowding ($>140$ patients) is $0.085\%\pm0.009\%$, compared to $0.077\%$ in the data, see inset of Fig.~\ref{fig:fig5}B. These probabilities are equivalent to $\sim 7$ hours per year. Remarkably, the accuracy of predicting such a rare event, occurring with probability $<10^{-3}$, is within 10\%.

In addition, our model provides excellent predictions for relative overcrowding. For example, the ratio of the probabilities of observing relative overcrowding greater than 40\% and 20\% is $0.086$ in the data, and  $0.074\pm0.01$ in the model,  within 15\% accuracy (see inset of Fig.~\ref{fig:fig5}C). 
%{\color{red}In addition, the ratio of probabilities to observe above than $120$ and above than $100$ patients ($100$ patients being an overcrowded state but not severe according to NEDOCS) is $0.166$ in the data, while the model predicts $0.161\pm0.002$}

%In addition, the probability of having relative overcrowding greater than, say, 40\% over the average ({\color{red}what's the percentage for severe overcrowding?}) is ??? in the data , which deviates by ??? from the model's prediction, see Fig.~\ref{fig:fig5}.C). 

\subsection*{Parameter Elasticity of Crowding}
An important contribution of our model lies in its ability to explore the elasticity of crowding to arrival flux and length of stay.
A change in the arrival flux $f(t)$ can potentially occur due to population growth or changes in the medical condition of the population in the surrounding area. While the mean-field description [Eq.~(\ref{DREsol})] indicates that the average number of patients, and the average patient hours depend linearly on the arrival flux $f(t)$, surprisingly, the probability of observing severe or dangerous overcrowding, $P(n\!>\!120)$, is highly non-linear in $f(t)$. As shown in Fig.~\ref{fig:fig6}A, while a 10\% increment in the arrival flux increases the average number of patients by only 10\%, the number of these extreme overcrowding events will increase by a factor of $2.8$.%, i.e, an increase of 180\%. 

A patient's average length of stay in the ED is given by $1/\beta$, and is a metric that could potentially be mitigated by the ED management through better allocation of staff, or more efficient organizational procedures. Figures~\ref{fig:fig6}B,C  show the change in the probability of overcrowding as a function of the change in $1/\beta$ (in minutes), and as a function of the change in $\sigma_2$ (the amplitude of systematic noise), respectively. We find that overcrowding in the ED is extremely sensitive to the length of stay; e.g., shortening the typical length of stay by just 20 minutes (7-10\% of the typical time, depending upon the day) reduces the number of severe and dangerous overcrowding events by 52\% (Fig.~\ref{fig:fig6}B). In addition, we find that the sensitivity to the length of stay is much higher than the sensitivity to the systematic noise; e.g., lowering the systematic noise amplitude by 15\% decreases the probability of a severe or dangerous overcrowding event by 22\% (Fig.~\ref{fig:fig6}C), whereas a similar decrease in this probability  can be acquired by lowering the typical length of stay by just 2.5-3\% (7 minutes). 

Such calculations enable the ED to better allocate investments in various crowding mitigation initiatives. Reducing the typical length of stay (e.g., by addressing process bottlenecks, changing staff allocation, speeding up test results) is, as per our results, more effective in mitigating severe or dangerous overcrowding than is reducing the systematic noise (e.g., by investing in maintenance and service contacts for equipment,  avoiding fluctuations due to equipment down-time,  better allocating critical equipment such as MRI scanners, or by cross functional training of staff to compensate for staff absenteeism).

\begin{figure}[h]
\centering
\includegraphics[width=0.90\linewidth]{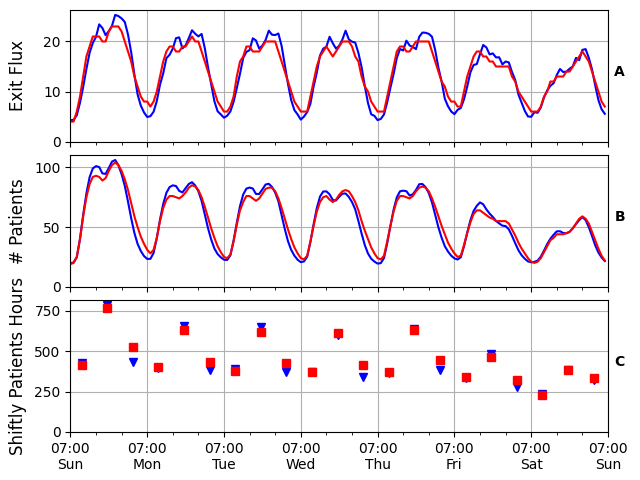}
\caption{Comparison between the data (blue) and the model (red) for the average exit flux (A), average momentary number of patients (B) and the average patient hours per shift (C).
Here $R^2$ equals $0.929$ (A) and $0.956$ (B)
}
\label{fig:fig4}
\end{figure}

\begin{figure}[htb!]
\centering
\includegraphics[width=0.90\linewidth]{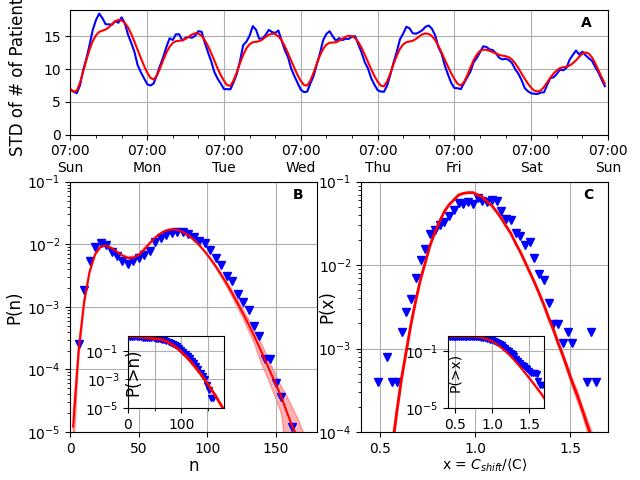}
\caption{ Model (red) fit with crowding statistics data (blue). (A) The fit of the hourly standard deviation of the number of patients. (B) Semi-logarithmic histogram of the number of patients in the ED in the 10 most crowded weekly shifts (Sunday through Thursday morning and afternoon). The shaded region accounts for the uncertainty in the estimation of the theoretical parameters. The inset shows the cumulative distribution. 
(C) Semi-logarithmic histogram of the patient hours of a shift relative to the average $x = C_{\text{shift}}/\langle C\rangle$ in the 10 most crowded weekly shifts. The inset shows the cumulative distribution. 
The $R^2$ value for (A) is $0.921$. The goodness-of-fit for (B) and (C), measured by the Kullback-Leibler divergence, is $0.02$ and $0.05$ respectively. }
\label{fig:fig5}

\end{figure}

\begin{figure}[htb!]
\centering
\includegraphics[width=0.90\linewidth]{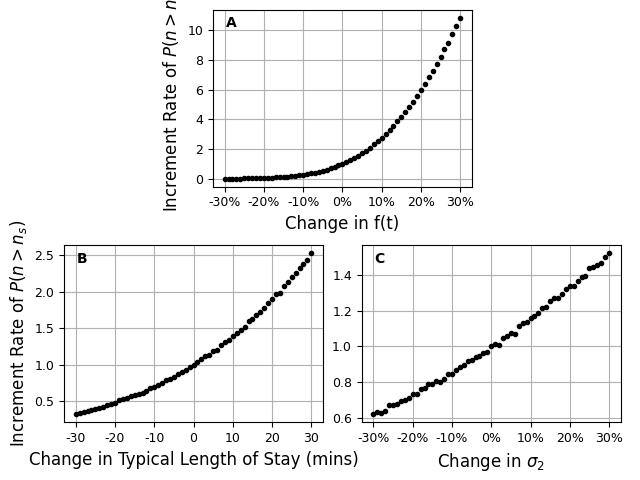}
\caption{Parameter Elasticity.
The increment rate of the probability for severe or dangerous overcrowding: the ratio of $P(n>n_s)$ to $P(n>n_s)$ at zero change, as a function of the change in the (A) arrival flux $f(t)$, (B) typical length of stay and (C) systematic noise magnitude $\sigma_2$.
Here $n_s = 120$ denotes the onset of severe overcrowding. Also, the average length of stay of the data is 3.9 hours, so approximately (depending on the weekday) a change in 10 minutes in the length of stay corresponds to 
4\%.
}
\label{fig:fig6}
\end{figure}

\section*{Discussion}
This paper addresses crowded human environments, which are characterized by high volatility; variation across hours and days; overcrowding events; inter-individual heterogeneity, and systematic noise. Modeling crowding in such a way as to account for all of these factors is critical for mitigating overcrowding and preventing service breakdowns. 

We present a theoretical framework using stochastic population modeling, an approach that has been applied thus far to describe a broad range of natural phenomena, but not on crowding problems. The model captures the arrival flux, the exit rate, and includes a combination of additive and multiplicative noise. We implemented the model on data from a hospital emergency department, and found that our model provides adequate prediction of the momentary number of patients, the standard deviation and the patient number distribution. Notably, the model accurately predicts the probability of exceeding a certain crowding threshold. More importantly, the model's main strength lies in its ability to predict how such overcrowding probabilities  vary if the model parameters are changed. This ability, absent from various ``black-box" models, is an important tool for the mitigation of overcrowding in the ED and the prevention of ED breakdowns. 

We used the model to explore several important "what-if" questions: how does population growth affect severe overcrowding in the ED? What happens if the length of stay is shortened by a certain amount? And: What is the effect of changing the systematic noise.  

%To demonstrate the model's novelty over that of existing crowding models, we compared its results to results obtained from a queuing theory model (see Appendix D in the SI), and we discovered that the queuing model required considerably more parameters (e.g., hour-by-hour staff average), yet its estimation was significantly less accurate than the model presented here, both in averages and in distribution estimations.

This work is important to both practice and research. ED practitioners can use the model to better allocate resources as per their expected effect. They can predict the volatility and rate of expected overcrowding events of any given magnitude. On the theoretical side, this work demonstrates the power of the stochastic population dynamics formalism, which has been widely used for describing various effects, e.g., in population biology, ecology, epidemiology, chemistry, and statistical physics. We show that this formalism can also be applied to describe the dynamics of populous human environments. The latter, while influenced by numerous variables and characterized by high volatility and overcrowding spikes, can be captured by a compact analytic formulation that enables revealing key underlying mechanisms, and obtaining important insights as to the role of various governing parameters. 

This research has several limitations: First, being a stochastic model, this approach cannot predict a specific date of an overcrowding event, but rather, it provides the distribution and  probability for such events to occur. Second, this study does not account for several low-frequency phenomena that can be found in the data, such as growth in demand over the years, or annual seasonality and systematic failures which may last several hours or days. Third, the model does not incorporate exogenous catastrophes such as large-scale accidents, extreme weather, or medical staff strikes. 

This work paves the way for future research in several directions: First, the model can be implemented on other populous environments (such as public transportation, shopping malls, etc.). Second, one can study the determinants of the model parameters (the dependence of exit rates on the total population in the ED, staff fatigue, medical team composition etc.). Another avenue could be expanding the model with other stochastic behaviours, such as low-frequency trends and noises, or an arrival flux with stochastic burst generators mimicking large accidents or other catastrophes.

\section*{Methods}

\subsection*{Data}

We use the complete set of records covering all ED visits between January 1, 2013 and June 30, 2018, from a large state-run Israeli hospital. Each record represents a visit by a single individual, and for each such record the dataset contains, inter alia, an encrypted patient identifier (to enable tracking of revisits), reason for visit, gender, age, mode of arrival (ambulance vs. self-arrivals), triage urgency, lab test timings and results, and all of the medical decisions for the patient. The data include time of arrival to and time of departure from the ED (discharge or hospitalization), as well as time stamps for each recorded operation; thus each visit-log contains a minute-by-minute description of the patient visit. The dataset consists of 679,762 visits by 333,471 unique patients, 46.4\% of whom were females, 24\% under age 18 and 23\% age 65 or older. The most common reasons for arrival are sickness (72\% of cases) and injury (16\%). The average length of stay in the ED is 3.9 hours, with a standard deviation (STD) of 2.6 hours. 
%We also have data on all the clinical staff who worked in the ED (even if called to the ED from another department), containing their ID, role, and their attendance system clock information on entering and leaving the ED. Staff data consist of 42,065 attendance records for 48 unique physicians, and 99,437 attendance records for 65 unique nurses. The staff data also contains  attendance records of the non-clinical support staff.    

We use the weekly hour (e.g., Sunday 8:00-9:00) as a basic time unit, due to the population's strong weekly cycle, with high typical crowding during weekdays (Sunday-Thursday, following the Israeli workweek), and lower crowding on the weekend (Friday-Saturday). To align with the ED shift structure, we grouped the hours, when needed, into morning  (7:00-15:00), afternoon  (15:00-23:00), and night  (23:00-7:00) shifts. 

\subsection*{Crowding Metrics}
Research and practice suggest various ways to measure ED crowding, differing in the data they require and the purpose of measurement. Some methods capture the inflow of patients (total number of daily visitors, current number of patients being treated or waiting to be seen), while others capture the load experienced by the patients or the staff (waiting times, treatment times, patients who leave without being seen, nurses being rushed or feeling rushed, etc.). A considerable number of measures focus on the facility's physical infrastructures (number of available beds, capacity in observation area, patients placed in ED hallways, etc.), see \cite{hwang2004care} for review.
A popular measure is the National Emergency Department Overcrowding Score (NEDOCS) \cite{weiss2004estimating}, officially used by the USA federal authorities. This score is a multi-variable function based on both site-specific parameters (total beds in the ED, Number of hospital beds), and momentary indices (total number of patients, average waiting time, etc.)

For this research, we focused on four metrics: arrival flux, exit flux, momentary number of patients, and ``patient hours''.
The arrival flux $f(t)$ is the hourly number of incoming patients (Fig.~\ref{fig:fig2}A).  The exit flux is the number of patients who left the ED per hour (for both discharge and hospitalization, see Fig.~\ref{fig:fig2}B).  The momentary number of patients is the number of patients who are currently in the ED, denoted by $n(t)$ (Fig.~\ref{fig:fig2}C). The metric of patient hours, which was specifically developed for this research, is defined by $C =\int_{T}^{T+\Delta T}n(t)dt$ (see grey area under the curve in Fig~\ref{fig:fig2}C and Fig.~\ref{fig:fig2}D). This is the accumulated number of patients in the ED during the time interval $[T,T+\Delta T]$, where T is measured in hours. 
That is, if the shift began with 100 patients, and they all remained throughout the shift, the patient hours for that shift will be $100*8=800$.
While the measure of patient hours, which is a combination of several other commonly used measures \cite{bellow2014evolution}, does not account for factors such as bed availability, waiting times, etc.,  its key advantage is that it is easier to integrate over shifts and days, and therefore, is more suitable for forecasting future ED occupancy.

The average hourly arrival flux of patients in our data is 14 (STD=8, min=0, max=46). The average hourly exit flux of patients is likewise 14 (STD=7, min=0, max=46). The fact that their average coincides naturally indicates that there is no long-term accumulation of patients. The average momentary number of patients present in the ED (as recorded hourly) in our data is 55 (STD=27, min=3, max=161), while the average patient hours per shift  is 440 (STD=164, min=130, max=1,198).

\subsection*{Crowding Over Time}
Figure~\ref{fig:fig2} shows strong daily and weekly cycles. Most of the patients arrive in daytime, and most of them, even those who arrived in the afternoon or evening, tend to leave before the late nighttime. During weekends, there are fewer patients in the ED than there are during the week.
Figure ~\ref{fig:fig2}D depicts the distribution of patient hours per shift. The figure clearly shows the daily/weekly cycles. The afternoon shift (15:00-23:00) is typically more crowded than are the morning or night shifts; and the workweek is more crowded than the weekend, with the first day of the workweek (Sunday) being the most crowded. 
Figure~\ref{fig:fig3}B shows, for each cohort of patients arriving at a given time $t_0$, the percentage of the remaining patients at $t>t_0$. The figure shows a clear exponential-like decay with a constant decay factor $\beta>0$. It is convenient henceforth to define the exit flux and exit rate as the hourly number and hourly fraction of patients who left the ED, respectively. Thus, $\beta(t)$ represents the exit rate, or the rate at which a patient exits the ED (discharged, hospitalized) within the next hour.
The higher $\beta(t)$ is, the faster the patient turnover, and therefore $\beta$ may be a good metric for the ED efficiency.

\subsection*{Fitting the Model}
We estimated the model described by Eq.~(\ref{Lang}) using a two stage process: we first estimated the deterministic part, namely  $\beta(t)$, and then the stochastic components $\sigma_1,\sigma_2$. 
We chose to extract $f(t)$ from the empirical data. We confirmed that other choices of $f(t)$, e.g., various polynomials of various orders, yield similar qualitative results (see SI, Appendix C).

The data indicate that patients' exit process from the ED follows a Poisson distribution (Fig.~\ref{fig:fig3}B). Thus, if the only process is the exit of patients, the mean number of patients would simply decay exponentially with an exponent $\beta$. We fitted $\beta$ for each weekday according to Eq.~(\ref{DREsol}). The $\beta$ values are given in Table~\ref{Param_table}. We chose four representative values for $\beta=\beta_i$: Sunday, midweek (Monday-Thursday), Friday, and Saturday.

The second stage entailed estimating the stochastic components $\sigma_1$ and $\sigma_2$. For every point in their two dimensional parameter space $\bm{\sigma} = (\sigma_1,\sigma_2)$, we ran $10^4$ realizations of a simulated week, calculating the hour-by-hour STD, denoted by $S(\bm{\sigma},t)$, and compared it to the data hourly STD, $S_i$. Here, $S_i = S_i(t_i)$ is comprised of $n$ points of time denoted by $t_i$ (measured in hours). We estimated $\bm{\sigma}$ by maximizing the Likelihood function:
\begin{equation}
\hspace{-2mm}\mathcal{L}(\bm{\sigma},p|\{t_i,S_i\}_{i\!=\!1}^n\!) \! =\!  (2\pi\sigma^2)^{-n/2}\prod_{i=1}^n\!\exp\!\!{\left[-\frac{(S_i\!-\!S(\bm{\sigma},t_i))^2}{2p^2}\right]}\!,
\end{equation} 
where we assumed that the sampled data have additional white Gaussian noise with variance $p^2$. The Gaussian assumption is justified as the hourly arrival flux exhibits a Poisson distribution, which, in the limit of large numbers, and especially in the right tail of the distribution, can be regarded as a Gaussian. 

By differentiating the log-likelihood function with respect to $\bm{\sigma}$ and $p^2$ and equating to zero, we find the maximum likelihood for the value of $\bm{\sigma}^*$ that minimizes $\sum_{i=1}^n(S_i-S(\bm{\sigma},t_i))$ -- the minimum of the mean square error (MSE). In addition, this procedure provides the value of $p^2$, which satisfies $p^2 = \text{MSE}(\bm{\sigma}^*)$.
The uncertainty in $\sigma_1$ and $\sigma_2$ is estimated by fitting for each $\sigma$ to a Gaussian:
\begin{equation}
   \mathcal{L}(\sigma_i) = C\exp{\left[-\frac{n}{2\,\text{MSE}(\bm{\sigma}^*)}\text{MSE}(\sigma_i)\right],}
\end{equation} 
with the parameter's uncertainty as its width, and $C$ being a constant.

\subsection*{Curve Fitting Statistics}
Table~\ref{Param_table} presents the values of the estimated parameters. Interestingly, although we did not fix the noise magnitude $\sigma_1$ to the value of $1$, to allow for additional sources of noise due to population heterogeneity, the maximum likelihood method estimated $\sigma_1\simeq 1.1$. This value is consistent with the Fokker-Planck approximation to the master equation, for which $\sigma_1 = 1$ (see SI, Appendix A) \cite{assaf2017wkb}. We confirmed that fixing $\sigma_1=1$ has a negligible effect on the results; the Kullback-Leibler divergence between the data and model patient number distribution, as displayed and calculated in Fig.~\ref{fig:fig5}B, changes from 0.02 to 0.029 in this case.

\begin{table}[ht]
\centering
\caption{Fitted parameter values for the model}
\begin{tabular}{lr}
Parameter &  Value  \\
\hline
$\beta^{Sun}$ & 0.224 \\
$\beta^{Mid}$ & 0.244  \\
$\beta^{Fri}$ & 0.280\\
$\beta^{Sat}$ & 0.304 \\
$\sigma_1$ & $1.1\pm0.015$ \\
$\sigma_2$ & $0.36\pm0.01$ \\
\hline
\end{tabular}
\label{Param_table}
\end{table}

\section*{Acknowledgments}
We would like to thank Efrat Naor, Michal Elchanan, Noam-Lee Kopivker and Yakov Lacher for their help with the data collection and early analysis. We thank the ED staff of the Shamir Medical center and Dr. Daniel Trotzky, the ED manager, for their invaluable advice and ongoing support. We thank Ohad Vilk for useful discussions.  RP acknowledges support from the ISF and the KMart foundation. MA acknowledges support from ISF grant 531/20.

\paragraph*{Author contributions:}
GP, RP and MA designed the model and performed the data analysis. OLK collected the data and contributed operational insights. GP, RP and MA performed the analytical and numerical calculations. GP, RP and MA wrote the manuscript.

\paragraph*{Competing interests:}
The authors declare that they have no competing interests.

% Your references go at the end of the main text, and before the
% figures.  For this document we've used BibTeX, the .bib file
% scibib.bib, and the .bst file Science.bst.  The package scicite.sty
% was included to format the reference numbers according to *Science*
% style.

\bibliography{scibib}

\bibliographystyle{Science}

\newpage
\onecolumn

\setcounter{figure}{0}
\renewcommand{\thefigure}{S\arabic{figure}}

\setcounter{table}{0}
\renewcommand{\thetable}{S\arabic{table}}

\renewcommand{\theequation}{S\arabic{equation}}    
\setcounter{equation}{0} 

\begin{center}\Huge{Supplementary Information}
\end{center}

\vspace{.5cm}

\normalsize
\section{Governing Equations for the Stochastic Model}

\subsection{Influx-Outflux Process}
The stochastic dynamics of incoming and outgoing patients consist of two stochastic processes: creation ("influx" or "arrival") and decay ("outflux" or "exit") of particles, which can be written as

\begin{equation}
    \emptyset \xrightarrow{f} A,  \qquad    A\xrightarrow{\beta}\emptyset.
    \label{rates}
\end{equation}
Note that the rate $f$ is absolute, and $\beta$ is per particle.
The dynamics described by this set of reactions is often used to model various chemical processes, or more generally, any Poisson process. %\cite{connors1990chemical}.
%Nevertheless, this set of reactions may be applied to describe influx and outflux of individuals.

\subsection{Deterministic Rate Equation}
At the deterministic level, the average number of particles $\Bar{n}$ as a function of time satisfies the following rate equation:
\begin{equation}
    \Dot{\Bar{n}} = f - \beta \Bar{n}(t).
    \label{DRE}
\end{equation}
This equation is obtained in a straightforward manner from rates~(\ref{rates}). The population growth rate is given by $f$, while the degradation rate per individual is given by $\beta$. 

The dynamics of Eq.~(\ref{DRE}) are simple, and can be solved analytically. The equation admits an attracting fixed point at $\Bar{n}_* = f/\beta$. Upon starting at any given $n_0 > 0$, the system will converge to the fixed point after a typical timescale on the order of $\tau = \beta^{-1}$. The exact solution reads:
\begin{equation}\label{DREsol1}
    \Bar{n}(t) = \bar{n}_* + (n_0-\bar{n}_*)e^{-\beta t}.
\end{equation}

Note that, the rate equation (here, and in general) describes the average behaviour of the system, and ignores demographic  fluctuations. This is justified as long as the typical population size satisfies  $N\gg 1$. Notably, Eqs.~(\ref{DRE}) and (\ref{DREsol1}) are the time-independent version of Eqs.~(\ref{DRE1}) and (\ref{DREsol}) in the main text, respectively. The numerical solution of the rate equation  for our case can be seen in Fig.~\ref{fig:fig4}.

To account for the intrinsic noise related to the discreteness of particles and stochasticity of the particles involved, a master equation can be used.

\subsection{Stochastic Case: the Master Equation}
The master equation is a gain-loss equation, describing the evolution of the probability $P_n(t)$ of observing n particles at a given time t, where n is discrete and t is taken to be continuous. 
% The generic form of the master equation is:
% \begin{equation}
%     \frac{dP_n(t)}{dt} = \sum_r\left[W(n,r)P_{n-r}(t) - W(n,r)P_n(t) \right],
%     \label{master_generic}
% \end{equation}
% where $W(n,r)$ is the rate at which the system jumps from a state with $n$ particles to a state with $n+r$, and terms that include $P_k$ with $k<0$ are assumed to be zero (no negative population).
% Applying Eq.~(\ref{master_generic}) on the simple model described above (\ref{rates}) yields:
For the stochastic process described by rates~(\ref{rates}),  the master equation yields:
\begin{equation}
    \begin{split}
    \frac{dP_n(t)}{dt} = f [P_{n-1}(t) - P_n(t)] + \beta [(n+1)P_{n+1}(t) - n P_n(t)].
    \end{split}
    \label{master}
\end{equation}
The solution of this equation, $\{P_n(t)\}_{n=0}^\infty$, for any n and t,  yields the probability distribution function (PDF) of the system, which describes typical fluctuations, as well as  rare events of interest. When the rates $f$ and $\beta$ are explicitly time dependent, an analytical solution of Eq.~(\ref{master}) is unknown in general. However, in the time-independent case, a solution can be found, e.g., using the method of characteristics~\cite{gardiner1985handbook}. 

Here we provide the stationary solution for the PDF of Eq.~(\ref{master}). Letting $\Dot{P}_n(t) = 0$ enables finding a recursive relationship between $P_n$ and $P_{n-1}$. After some algebra, we obtain:
\begin{equation}\label{statsol}
    P_n =  P_0 \prod_{k=0}^{n-1} \frac{f}{\beta(k+1)} = P_0 \left(\frac{f}{\beta}\right)^n\frac{1}{n!},
\end{equation}
where $P_0$ is found from the normalization condition: $\sum_{n=0}^{\infty} P_n = 1$. This PDF is simply  a Poisson distribution with mean $\lambda = f/\beta$, and therefore, one finds $P_0 = e^{-\lambda} = e^{-f/\beta}$.

Notably, the rate equation can be directly obtained from the master equation~(\ref{master}). Multiplying Eq.~(\ref{master}) by n, summing over all $n$'s, and using the relation $\Bar{n} = \sum_nnP_n$, we obtain Eq.~(\ref{DRE}) upon neglecting subleading-order terms and assuming $N\gg1$.

\subsection{The Fokker-Planck and Langevin Equations for the Influx-Outflux Process}
The Fokker-Planck equation (FPE) or Langevin equation are best known from applications of classical
mechanics, and describe the dynamics of a particle moving in a deterministic force field with some noise. 

\begin{figure}[ht]
\centering
\includegraphics[width=.7\linewidth]{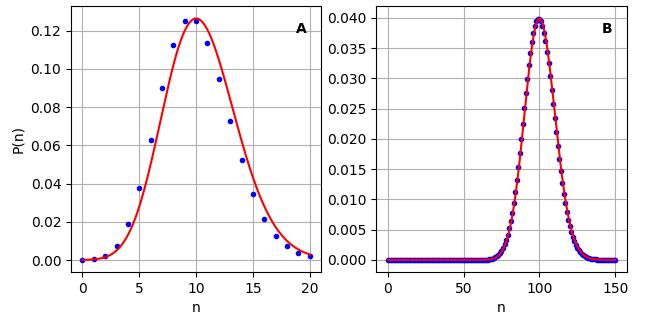}
\caption{Comparison between the solution of the master equation [Eq.~(\ref{statsol})] (blue dots) and that of the Fokker Planck Equation [Eq.~(\ref{FPEsol})] (red line) for 2 different cases
(A) $f=5$, $\beta = 0.5$, $n_* = f/\beta = 10$
(B) $f=25$, $\beta = 0.25$, $n_* = f/\beta = 100$.}
\label{fig:figS1}
\end{figure}

When the starting point is a master equation, the FPE can be obtained using the  van-Kampen
system size expansion~\cite{gardiner1985handbook,assaf2017wkb}, valid for large $n$. In the case of the influx-outflux process~(\ref{rates}), the FPE yields:
\begin{equation}
\begin{split}
    \frac{\partial P(n,t)}{\partial t} =  -\frac{\partial}{\partial n}\left[P(n,t)(f-\beta n)\right] + \frac{1}{2}\frac{\partial^2}{\partial n^2}\left[P(n,t)(f+\beta n)\right].
\end{split}
    \label{fpe_solved}
\end{equation}
In the time-independent case, the stationary PDF is found by equating the left-hand side of Eq.~(\ref{fpe_solved}) to zero, and using the zero-flux boundary condition:
\begin{equation}\label{FPEsol}
    P(n) = C\,\exp{\left(2\int_{0}^n \frac{f-\beta n'}{f + \beta n'}dn'\right)},
\end{equation}
where C is a normalization factor, found by demanding that the PDF be normalized to 1. To test the quality of the Fokker-Planck approximation, we compare in Fig.~\ref{fig:figS1} the solution of the master equation to that of the FPE. As expected, as the typical value of $n$ increases, the accuracy of the FPE improves.

Notably, the equivalent Langevin equation to the FPE, given by Eq.~(\ref{fpe_solved}), reads:
\begin{equation}
    \frac{dn(t)}{dt} = f - \beta n(t) + \sqrt{f+\beta n(t)} \eta(t).
    \label{SDE}
\end{equation}
This equation describes the randomly-varying number of patients subject to deterministic forcing, $f-\beta n$, and multiplicative noise, $\sqrt{f+\beta n(t)} \eta(t)$, where $\eta(t)$ is delta-correlated Gauassian noise.

Equation~(\ref{SDE}) is similar to Langevin equation in the main text [Eq.~(\ref{Lang})], with time-independent rates, $\sigma_1=1$, and without systematic noise ($\sigma_2 = 0$). Note that, in the case of explicitly time-dependent rates, in general the PDF cannot be found analytically, and one has to resort to numerical techniques.

%\newpage

\section{Accounting for Colored Noise}\label{secA3}
Here we consider the scenario where both the internal and external noise have a finite correlation time. Consequently, instead of taking delta-correlated noise, we complement the Langevin equation for the fluctuating patients number [Eq.~(\ref{Lang}) in the main text] with two Ornstein-Uhlenbeck equations:
\begin{equation}
    \frac{d\xi_1(t)}{dt} = \frac{\xi_1(t)}{\tau_1} + \sqrt{\frac{2\sigma_1^2}{\tau_1}}\zeta_1(t)\quad\quad\quad
    \frac{d\xi_2(t)}{dt} = \frac{\xi_2(t)}{\tau_2} + \sqrt{\frac{2\sigma_2^2}{\tau_2}}\zeta_2(t),
\end{equation}
where $\sigma_i$ are the amplitudes of the noise, $\tau_i$ are the noise correlation times, and $\zeta_i$ are  delta-correlated Gaussian noise terms.

To study these scenarios, we used the same maximum likelihood estimation method (as described in Methods section), and fitted $\sigma_1,\sigma_2$ for each pair of noise correlation times, see Table~\ref{KL_Table}. To do so, we calculated the Kullback-Leibler divergence, for the same case of the 10 most crowded shifts of the week (as done in Fig.~\ref{fig:fig5}B), see Table~\ref{KL_Table}. The results do not vary much from the value of $0.02$ that we obtained for the delta-correlated case (which corresponds to the case of $\tau_1,\tau_2 = 1$, the shortest correlation time possible for this dataset), and range from 0.016 to 0.032. This implies that while our choice of delta-correlated noise was arbitrary, accounting for finite correlation time of the noise does not introduce notable changes to the model's predictions. 

\begin{table}[htb!]
        \centering
        \caption{Kullback–Leibler (KL) divergence between the data and model for the patient-number distribution of the 10 most busy shifts, for correlated noise (Uhlenbeck-Ornstein noise) with various correlation times. Here $\tau_1$ differs across columns (2-6) while $\tau_2$ differs across rows.\\ }
        \begin{tabular}{lrrrrr}
  $\tau_2\quad$ & $\tau_1=1$ &  2 & 4 & 8 & 24  \\
\hline
%        \headercell{Noise correlation  times} & \multicolumn{5}{c@{}}{$\tau_1$} %\\
%        %\cmidrule{2-6} 
%        \\ $\tau_2$ & 1 &  2 & 4 & 8 & 24   \\ 
%        \hline
        1  & \quad 0.020 &  0.021 &  0.019 &  0.022 &  0.021  \\
        2  & 0.032 &  0.020 &  0.018 &  0.016 &  0.017  \\
        4  & 0.031 &  0.020 &  0.018 &  0.018 &  0.017 \\
        8 & 0.031 &  0.019 &  0.018 &  0.019 &  0.016 \\
        24 & 0.029 &  0.018 &  0.018 &  0.019 &  0.019  \\
       
    \end{tabular}
    \label{KL_Table}
\end{table}

%\newpage
\section{Alternative Models For the Arrival Flux}\label{secA2}
\begin{figure}[t]
\centering
\includegraphics[width=.75\linewidth]{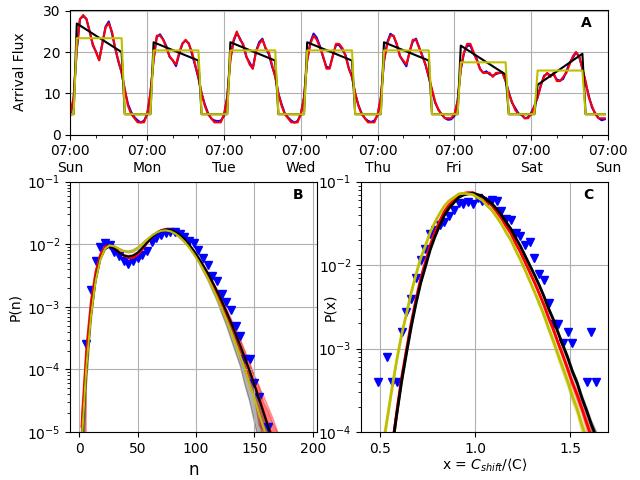}
\caption{Crowding statistics of the data (blue triangles) fit with three arrival flux models (measured arrival flux in red, trapezoid in black, rectangular in yellow). (A) The fit of the hourly arrival flux. (B) Semi-logarithmic histogram of the number of patients in the ED in the 10 most crowded weekly shifts (Sunday through Thursday morning and afternoon). The shaded region accounts for the uncertainty in the estimation of the theoretical parameters.
(C) Semi-logarithmic histogram of the patient hours of a shift relative to the average $x = C_{\text{shift}}/\langle C\rangle$ in the 10 most crowded weekly shifts. 
}
\label{fig:figS2}
\end{figure}

We implemented several other population dynamics models based on the Langevin equation, using several other arrival fluxes. For example, we implemented the model described by Eq.~(\ref{Lang}) using a simple trapezoid function of the form:

\begin{equation}
    f(t) = \left\{ \begin{array}{cc} a_1^i+ \frac{a_2^i-a_1^i}{t_2-t_1}(t-t_1) & \quad t_1 < t < t_2 \\ b & \quad else \end{array}\right. ,
    \label{trap_flux}
\end{equation}
where $a_1^i,a_2^i>b$ are the rates at the rush hours, with $i$ (here and below) indicating various parts of the week ($i\in\left\{\mbox{Sunday, Midweek, Friday, Saturday}\right\}$); and $b$ is the arrival flux at the off-peak hours (mostly during the night or early morning).  $\beta(t)$ differs between the various parts of the week, such that $\beta$ is taken to be $\beta^i$. 
Notably, We also tested a simpler, rectangular arrival flux, which has a constant value during rush hours: $a_1^i = a_2^i$ for every part of the week. The fitted arrival flux functions can be seen in  Fig.~\ref{fig:figS2}A. For each of the models we repeated the process of fitting the stochastic components. The fitted parameters can be seen in Table.~\ref{Param_table_full}

Using the various models we estimated  the patient number distributions for every model. Figures \ref{fig:figS2}B and \ref{fig:figS2}C show  excellent agreement between the crowding statistics of the models and data. This indicates that using simplified approximate functional forms of the arrival flux allows for a semi-analytical calculation of the PDF, as well as parameter elasticities, as done numerically in the main text.

\begin{figure}[t]
\centering
\includegraphics[width=.75\linewidth]{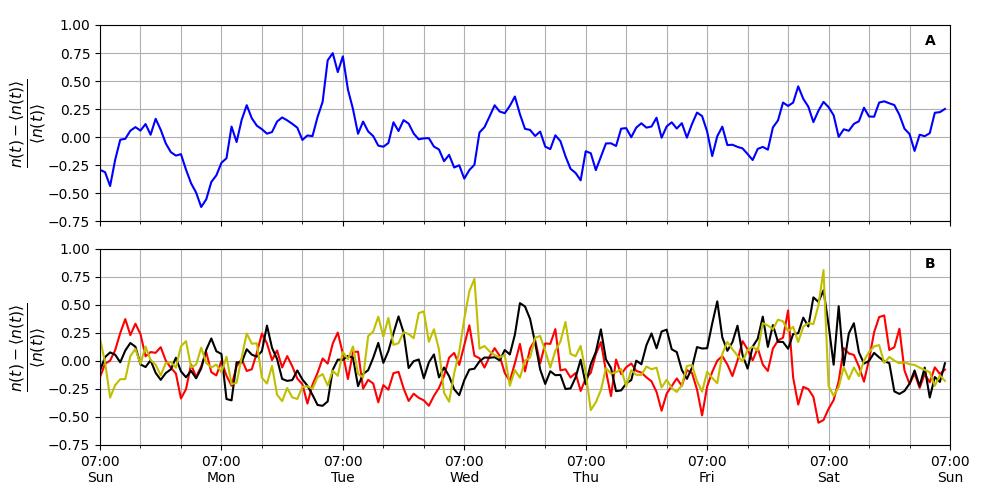}
\caption{Qualitative comparison of the data and model.
(A) The normalized deviation in the total number of patients relative to the average, $(n(t)-\langle n(t) \rangle)/\langle n(t) \rangle)$, in a typical week in the data.
(B) The normalized deviation in the total number of patients relative to the average in a typical realization of a week by our simulation using the various arrival flux models (measured in red, trapezoid in black, rectangular in yellow).}
\label{fig:figS3}
\end{figure}

Finally, Fig.~\ref{fig:figS3} illustrates the relative fluctuations in the total number of patients with respect to the hourly average, $(n(t)-\langle n(t) \rangle)/\langle n(t) \rangle$ over a typical week, for the data and the various models (see also Fig.~\ref{fig:fig1} in the main text). The figure shows that the relative fluctuations over the week can be very high, reaching 75\% above and 60\% below the average number of patients over just a few days. One can also see that it takes several hours for the ED to revert to the average crowding levels. This stresses the importance of accounting for noise  while  modeling crowding in such environments.

\begin{table}[ht]
\centering
\caption{Fitted parameter values, for the measured arrival flux (A), trapezoid arrival-flux model (B),  and rectangular arrival-flux model (C), used in the main text. The latter resembles model (B) [Eq.~(\ref{trap_flux})], but with a constant value during rush hours: $a^i_1 = a^i_2$ for every part of the week.}\vspace{1mm}
\begin{tabular}{lrrr}
Parameter & Measured arrival flux & Trapezoid arrival flux &  Rectangular arrival flux  \\
\hline
$t_1$ (hour) & - & 9:00 &  9:00 \\
$t_2$ (hour) & - & 23:00 &  23:00  \\
b & - & 4.98 & 4.98 \\
$a_1^{Sun}$ & - & 26.94 & 23.38 \\
$a_1^{Mid}$ & - & 22.39 & 20.42  \\
$a_1^{Fri}$ & - & 21.58 & 17.54 \\
$a_1^{Sat}$ & - & 12.1 & 15.55  \\
$a_2^{Sun}$ & - & 20.02 & - \\
$a_2^{Mid}$ & - & 17.96 & - \\
$a_2^{Fri}$ & - & 14.59 & - \\
$a_2^{Sat}$ & - & 19.61  & - \\
$\beta^{Sun}$ & 0.224 & 0.220 & 0.225 \\
$\beta^{Mid}$ & 0.244 & 0.241 & 0.245  \\
$\beta^{Fri}$ & 0.280 & 0.282 & 0.271 \\
$\beta^{Sat}$ & 0.304 & 0.295 & 0.271 \\
$\sigma_1$ & $1.10\pm0.015$ & $1.12\pm0.02$ & $1.20\pm0.02$ \\
$\sigma_2$ & $0.36\pm0.01$ & $0.350\pm0.005$ & $0.315\pm0.01$ \\
\hline
\end{tabular}
\label{Param_table_full}
\end{table}

\end{document}